\documentstyle[PASJadd,psbox]{PASJ95}

\newcommand{\lett}{}

\begin{document}
\title{Discovery of 101-s Pulsations from AX~J0057.4$-$7325 in the SMC 
with ASCA
}
\author{Jun {\sc Yokogawa},$^1$
Ken'ichi {\sc Torii},$^2$
Takayoshi {\sc Kohmura},$^3$
Kensuke {\sc Imanishi},$^1$
and Katsuji {\sc Koyama}$^1$\thanks{CREST, 
Japan Science and Technology Corporation (JST), 
4-1-8 Honmachi, Kawaguchi, Saitama, 332-0012.} \\ [12pt]
$^1$ {\it Department of Physics, Graduate School of Science, Kyoto University, 
Sakyo-ku, Kyoto, 606-8502} \\
{\it E-mail(JY): jun@cr.scphys.kyoto-u.ac.jp} \\
$^2$ {\it National Space Development Agency of Japan, 
2-1-1 Sengen, Tsukuba, Ibaraki, 305-8505} \\
$^3$ {\it Department of Earth and Space Science, Graduate School of Science, 
Osaka University, }\\
{\it 1-1 Machikaneyama-cho, Toyonaka, Osaka, 560-0043}}

\abst{The results from two ASCA observations of 
AX~J0057.4$-$7325 = RX~J0057.3$-$7325 
are presented. 
Coherent pulsations with a barycentric period of $101.45 \pm 0.07$~s 
were discovered in the second observation. 
The X-ray spectrum was found to be hard (photon index $\sim 0.9$) 
and unchanged through these observations, except for the flux. 
The ROSAT archival data show 
that AX~J0057.4$-$7325 exhibits a flux variation with a factor $\gtsim 10$. 
A discussion on a possible optical counterpart is given. 
}

\kword{pulsars: individual (AX~J0057.4$-$7325) --- 
stars: neutron --- X-rays: stars}

\maketitle
\thispagestyle{headings}

\section{Introduction}
Recently, many X-ray pulsars have been discovered 
in the Small Magellanic Cloud (SMC), and now 
$\sim 20$ X-ray pulsars are known in this galaxy. 
Most of them are interpreted as being X-ray binary pulsars (XBPs), 
because of their hard X-ray spectrum, long pulse period ($\gtsim 1$~s), 
large flux variability, 
and/or the existence of a massive star counterpart 
(Yokogawa et al.\ 2000 and references therein). 
Massive star counterparts have been mostly revealed to be 
a Be star by detailed optical observations on individual objects 
(Hughes, Smith 1994; Southwell, Charles 1996; 
Buckley et al.\ 1997; Cowley et al.\ 1997; 
Israel et al.\ 1997; Coe et al.\ 1998; 
Lamb et al.\ 1999; Stevens et al.\ 1999; 
Coe, Orosz 2000; Coe et al.\ 2000), 
except for SMC X-1 with a B (not Be) supergiant star 
companion 
(Bildsten et al.\ 1997). 

Haberl and Sasaki (2000) investigated 
the spatial correlation between SMC X-ray sources 
and emission-line objects (hereafter ELOs) cataloged by 
Meyssonnier and Azzopardi (1993; MA93) as well as 
Murphy and Bessell (2000). 
They found that most of the already established Be/X-ray binaries 
have a counterpart in these ELO catalogs, 
and thus proposed that other X-ray sources having an ELO counterpart 
are also likely to be Be/X-ray binaries. 
They concluded that the number ratio of Be/X-ray binaries to 
OB supergiant X-ray binaries in the SMC is extremely higher than that 
in our Galaxy. 
They also pointed out that 
the main body of the SMC contains no OB supergiant X-ray binary, 
suggesting a different star-formation history 
between the main body and the eastern wing. 
In this context, it would be important 
to search for pulsations from X-ray sources 
with no ELO counterpart.

In this letter, we report on the discovery of coherent pulsations with ASCA 
from AX~J0057.4$-$7325 (Torii et al.\ 2000), 
for which neither a Be star nor an ELO counterpart has been known. 
We also investigated several catalogs of optical stars 
to find a possible optical counterpart.

\section{Observations and Data Reduction}
Two ASCA observations pointed at the edge of the SMC main body 
have covered AX~J0057.4$-$7325. 
The observation dates were 
51309.583--51310.698 (hereafter, obs.\ A1) and 
51659.506--51660.598 (obs.\ A2), 
in unit of Modified Julian Day (MJD). 

ASCA carries four XRTs (X-ray Telescopes, Serlemitsos et al.\ 1995) 
with two GISs (Gas Imaging Spectrometers, Ohashi et al.\ 1996) and 
two SISs (Solid-state Imaging Spectrometers, Burke et al.\ 1994)
on each focal plane. 
In both observations, 
the GIS was operated in the normal PH mode 
with a time resolution of 0.0625/0.5~s, 
while 
the SIS was operated in the complementary 2-CCD mode 
with a time resolution of 8~s. 
The SIS data format in each observation was 
Faint/Faint with a level discrimination of 0.7~keV (A1) 
and 
Faint/Bright with a level discrimination of 0.55~keV (A2).

We first rejected any data obtained in the South Atlantic Anomaly, 
or in low cut-off rigidity regions ($<4$~GV), 
or when the elevation angle of the target from 
the earth's rim was less than $5^\circ$. 
We also rejected the SIS data obtained when the elevation angle 
from the bright earth was lower than $25^\circ$.
Particle events for GIS were removed by a rise-time discrimination method.
Hot and flickering pixels of SIS were rejected. 
After screening, the total available exposure times were 
$\sim 41$~ks (A1; GIS), $\sim 37$~ks (A1; SIS), 
$\sim 26$~ks (A2; GIS), and $\sim 17$~ks (A2; SIS). 
Since SIS has been severely damaged by particle radiation, 
the pixel-to-pixel fluctuation of the zero level has become 
significantly large. 
To compensate for the degradation, we made an RDD correction for obs.\ A1 
(Dotani et al.\ 1997, ASCA News 5, 14), 
which is reliably applicable only to the Faint mode data. 
We did not apply an RDD correction for obs.\ A2, 
because we simultaneously used 
the Faint and Bright data to obtain better statistics.

\section{Results}
\subsection{Source Identification}
In obs.\ A2, a source was detected at the center of the SIS chip, S1C1. 
Using the SIS image, we determined the source position 
using a method described by Ueda et al.\ (1999) 
to be
($00^{\rm h}57^{\rm m}29^{\rm s}\hspace{-2pt}.4$, $-73^\circ25'19''$) 
for equinox 2000, 
with an error radius of $30''$. 
We thus designate this source as AX~J0057.4$-$7325.
Positions determined with GIS in both observations 
or SIS in obs.\ A1 were consistent with the above coordinates.

We investigated several X-ray and optical catalogs 
to find counterparts of AX~J0057.4$-$7325. 
A ROSAT source, RX~J0057.3$-$7325 
(Haberl et al.\ 2000), and an optical star, 
MACS~J0057$-$734\#010 (Tucholke et al. 1996), 
were found within the ASCA error circle.

\subsection{Timing Analyses}
In each observation, 
we collected source photons from a circle 
with a radius of $\sim 3'\hspace{-2pt}.5$ 
centered on AX~J0057.4$-$7325. 
We made a barycentric correction on the photon arrival times 
and performed an FFT (Fast Fourier Transformation) 
on the event lists in each observation. 
From the power spectrum of the GIS data in obs.\ A2, 
we detected a significant peak at a frequency of $\sim 0.0099$~Hz 
(figure 1), 
which was confirmed with the SIS data. 
The maximum power of 49.6 was obtained with photons in the energy 
band of 1.1--7.6~keV. 
Because random fluctuation can cause such a large power 
with a very low probability of $2\times10^{-6}$, pulse detection 
is highly significant. 
We performed epoch folding on the GIS+SIS data, and determined 
the barycentric period to be $P = 101.45 \pm 0.07$~s. 

We also detected weak evidence for coherent pulsations from obs.\ A1. 
A maximum power of 34.1 was detected at $\sim 0.0098$~Hz 
with the GIS+SIS data. 
The probability to obtain such a power at any frequency from random events 
is not very small, 0.03\%, 
but this is a rather conservative estimation 
because the period search could be restricted 
to a narrow range of around $\sim 101$~s.
We thus performed an epoch folding search on the data, and determined 
the barycentric period to be $P = 101.47 \pm 0.06$~s. 

Figure 2 shows the pulse profiles of GIS+SIS in the energy bands 
0.7--2.0~keV and 2.0--7.0~keV for both observations.  
The pulse shape is broad, and there may be an additional sub-peak 
in the low-energy band. 
The pulsed fraction, defined as (pulsed flux)/(total flux) 
after removing the background, 
is $\sim 46$\% (A1) or $\sim 35$\% (A2) in the 2.0--7.0~keV band.


\begin{figure}
\hspace*{8mm} \psbox[xsize=0.4\textwidth]{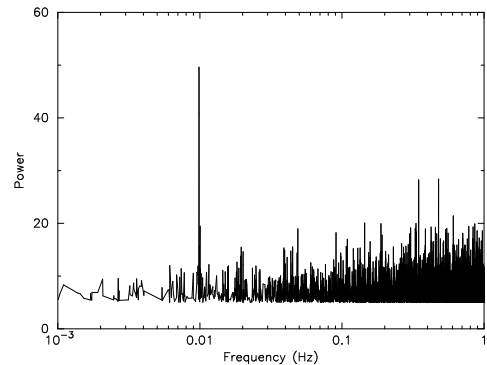}
\caption{Power spectrum obtained with the GIS data in obs.\ A2.  
Power is normalized so that random fluctuation level corresponds to 2. 
Data points with power less than 5 are omitted. 
An evident peak is detected at $\sim 0.0099$~Hz.}
\end{figure}

\begin{figure}[th]
\hspace*{16mm} \psbox[xsize=0.3\textwidth]{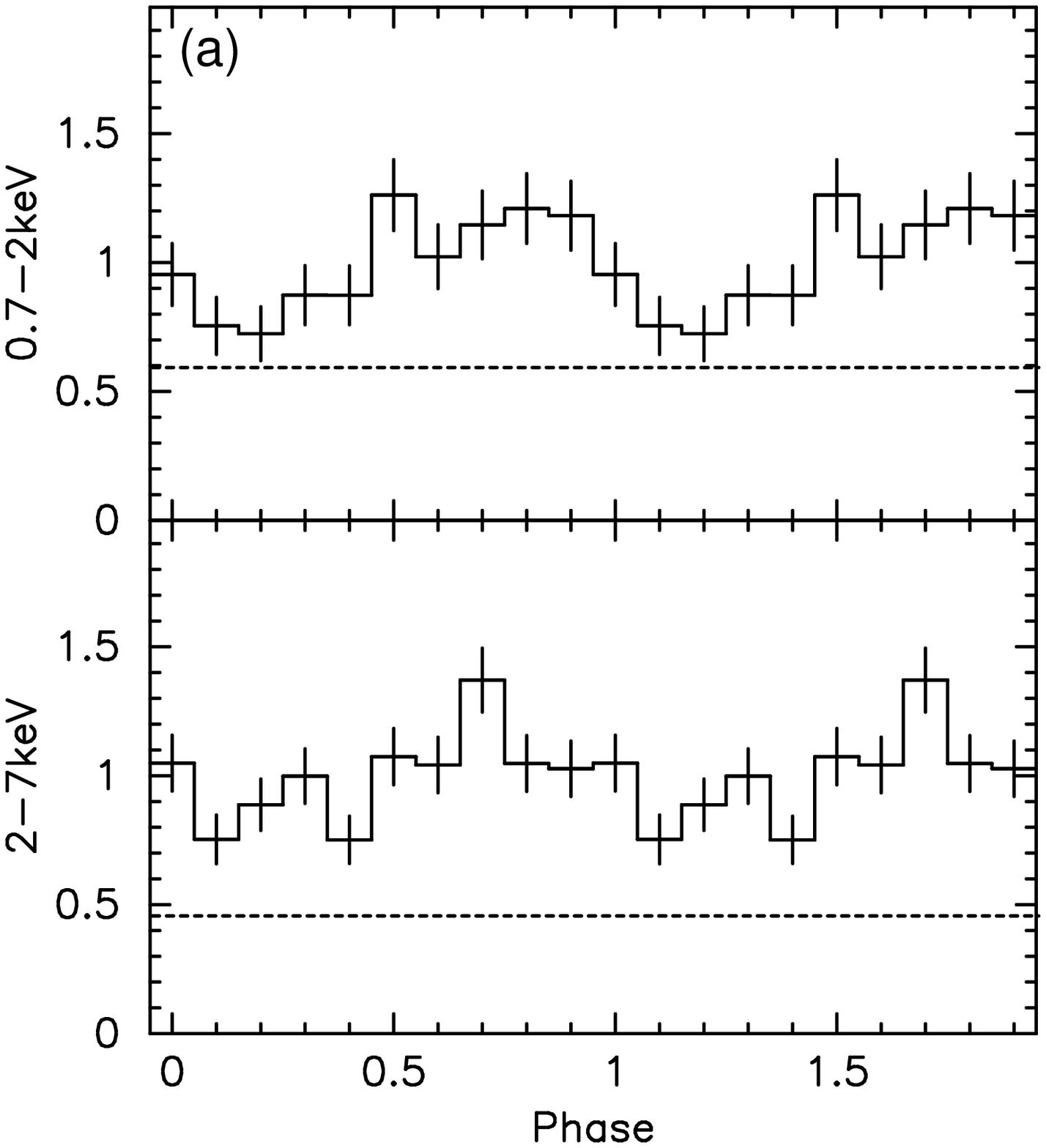}

\hspace*{16mm} \psbox[xsize=0.3\textwidth]{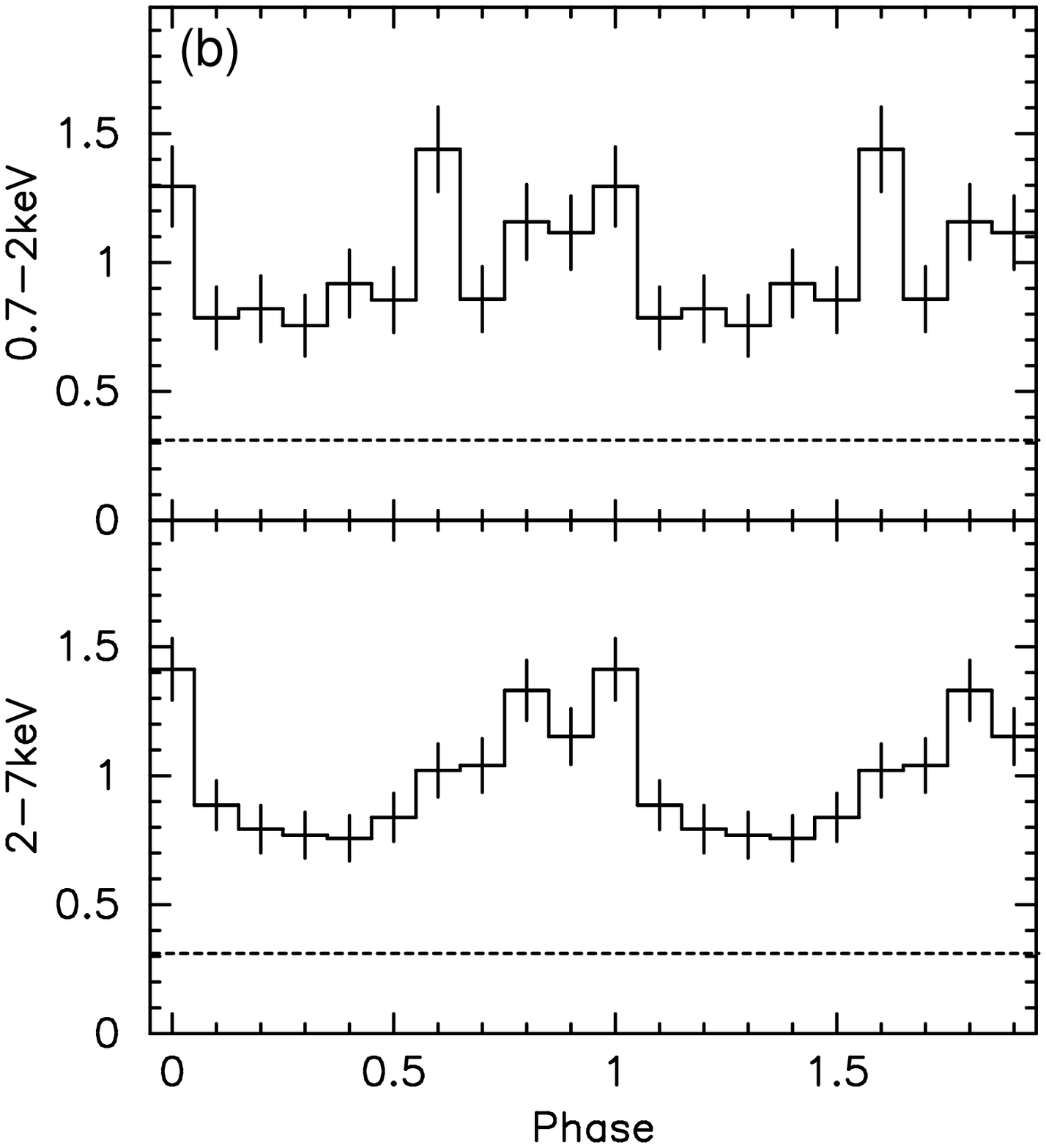}
\caption{Pulse profiles in obs.\ A1 (a) and A2 (b), where phase zero is arbitrary. 
Upper and lower panels are for 
the 0.7--2.0~keV and 
2.0--7.0~keV bands, respectively.
Vertical axes indicate normalized count rates in each energy band. 
Background levels are indicated by broken lines.}
\end{figure}

We also searched for an aperiodic intensity variation 
during each observation, but neither a burst nor a large flux variation 
was detected.

\subsection{Spectral Analyses}
Source photons were collected from the same regions used in 
the timing analyses, while background photons were 
from off-source areas near the source. 
We used the GIS and SIS spectra of obs.\ A1 and only 
the GIS spectrum of obs.\ A2. 
The SIS spectrum of obs.\ A2 was not used because 
we do not have a reliable response matrix for the Bright mode data, 
which is severely affected by radiation damage.

At first, we separately fitted each spectrum 
to a simple power-law model with the interstellar absorption. 
The derived parameters (photon index $\Gamma$ and column density $N_{\rm H}$) 
were found to be consistent in both observations and in both detectors. 
Therefore, 
to further constrain the parameters, 
we assumed that $\Gamma$ and $N_{\rm H}$ were the same 
for both observations, and simultaneously fitted the three spectra 
to the same model. 
We obtained the best-fit parameters as 
$\Gamma = 0.9$ (0.7--1.0) and $N_{\rm H} = 2$ (0.2--$5) \times 10^{21}$ 
cm$^{-2}$, with a reduced $\chi^2$ of 1.00 for 96 degrees of freedom
(the values in parentheses indicate 90\% confidence limits). 
X-ray fluxes were derived to be 
$1.2 \times 10^{-12}$ erg~s$^{-1}$~cm$^{-2}$ (for obs.\ A1) and 
$2.4 \times 10^{-12}$ erg~s$^{-1}$~cm$^{-2}$ (for obs.\ A2). 
Figure 3 shows the phase-averaged spectrum of GIS 
in obs.\ A2, which had the best statistics, 
and the best-fit model.

\begin{figure}[th]
\hspace*{8mm} \psbox[xsize=0.4\textwidth]{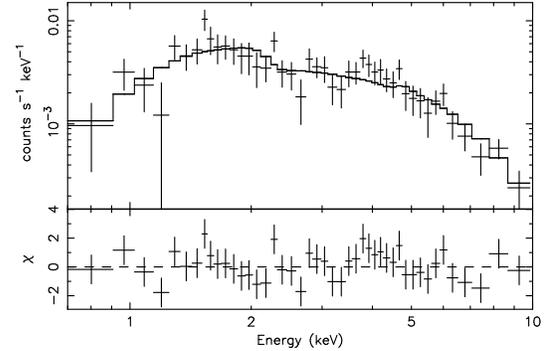}
\caption{Background-subtracted and phase-averaged spectrum in obs.\ A2 
(GIS 2+3). 
Crosses and a solid line indicate 
data points and the best-fit model, respectively. 
}
\end{figure}

We also extracted phase-resolved spectra from 
phases 0--0.5 and 0.5--1 (see figure 2) and fitted them 
with the same model. 
No significant difference was found for $\Gamma$ and $N_{\rm H}$ 
within the statistical errors, 
which is consistent with the energy-resolved pulse shapes (figure 2).

\section{Discussion}
Six ROSAT observations 
and no Einstein observation 
have covered the position of AX~J0057.4$-$7325. 
To investigate the long-term flux variation of this source, 
we used the ROSAT archival data as follows. 
We derived the count rate from AX~J0057.4$-$7325 
(= RX~J0057.3$-$7325) 
in each observation, 
after background subtraction. 
We assumed that $\Gamma$ and $N_{\rm H}$ were the same as 
those derived from the ASCA observations 
($\Gamma = 0.9$ and $N_{\rm H} = 2\times 10^{21}$ cm$^{-2}$), 
and converted the count rate to flux with the {\tt PIMMS} software. 
We also made a vignetting correction according to 
the source's off-axis angle in each observation. 
The results are given in table 1: 
AX~J0057.4$-$7325 was detected 
in observations R1 and R2 with a signal-to-noise ratio of 2 and 10, 
respectively. 
By comparing the fluxes in observations R4 and A2, 
we conclude that the flux has changed by a factor of $\gtsim 10$. 
The flux variability, the hard X-ray spectrum, and 
the long pulse period are all consistent with a scenario that 
AX~J0057.4$-$7325 is an XBP with a companion of 
either a Be, an OB supergiant, or a low-mass star.

As far as we have investigated, 
we found only one optical source, 
MACS~J0057$-$734\#010, in the ASCA error circle. 
It is interesting that 
there exists no ELO, 
i.e., a Be star candidate, 
in the ASCA error circle. 
Because the nearest ELO, No.\ 986 in MA93, is located $1.\hspace{-2pt}'4$
away from AX~J0057.4$-$7325, 
a possibility for the optical counterpart 
would be ruled out. 
This is a rare case 
where an X-ray pulsar in the SMC 
is not associated with a Be star or an ELO. 

{\small
\begin{center}
Table~1.\hspace{4pt}Flux variability of AX~J0057.4$-$7325. \\
\end{center}
\begin{tabular*}{0.5\textwidth}{@{\hspace{\tabcolsep}
\extracolsep{\fill}}p{2.5pc}ccc}
\hline\hline\\ [-6pt]
Obs.ID.&Date$^*$  &  Instrument  &  Flux$^\dagger$         \\
       &(MJD)     &              &  (erg~s$^{-1}$~cm$^{-2}$)      \\[4pt]\hline\\[-6pt]
R1\dotfill& 48550.301 & ROSAT/PSPC   & $1.1\times10^{-12}$ \\
R2\dotfill& 48732.665 & ROSAT/PSPC   & $1.3\times10^{-12}$ \\
R3\dotfill& 49118.070 & ROSAT/PSPC   &$<3.5\times10^{-13}$ \\
R4\dotfill& 49298.538 & ROSAT/PSPC   &$<2.4\times10^{-13}$ \\
R5\dotfill& 49321.271 & ROSAT/PSPC   &$<3.9\times10^{-13}$ \\
R6\dotfill& 49470.832 & ROSAT/HRI    &$<1.1\times10^{-12}$ \\
A1\dotfill& 51310.141 & ASCA/GIS+SIS & $1.2\times10^{-12}$ \\
A2\dotfill& 51660.052 & ASCA/GIS     & $2.4\times10^{-12}$ \\[4pt]\hline
\end{tabular*}
\vspace{1pt}\par\noindent
$*$ Middle of the observations.
\par\noindent
$\dagger$ In the 0.7--10.0~keV band, after vignetting correction.
}

We note that AX~J0057.4$-$7325 is located at the edge of 
the main body, fronting to the eastern wing. 
The fact that OB supergiant X-ray binaries in the SMC 
(only SMC X-1 and EXO 0114.6$-$7361) are both located in the eastern wing
may lead us to suspect that 
AX~J0057.4$-$7325 would be the third example. 
Because the discovery of a true optical counterpart would provide 
essential information, 
we strongly encourage deep and detailed optical observations 
on this field. 
\par
\vspace{1pc}\par
We thank Dr.\ M.J.\ Coe for his valuable suggestions 
about MACS~J0057$-$734\#010 and optical counterparts of the SMC sources. 
We also thank Dr.\ F.\ Haberl for useful comments on the SMC PSPC catalog.
The ROSAT data were obtained through the High Energy
Astrophysics Science Archive Research Center Online Service,
provided by the NASA/Goddard Space Flight Center. 
J.Y.\ and T.K.\ are 
supported by JSPS Research Fellowship for Young Scientists.
We are grateful for constructive comments from an anonymous referee. 

\section*{References} 
\small
\re
 Bildsten L., Chakrabarty D., Chiu J., Finger M.H., Koh D.T., 
 Nelson R.W., Prince T.A., Rubin B.C.\ et al.\ 1997, ApJS 113, 367
\re
 Buckley D.A.H., Coe M.J., Stevens J.B., Angelini L.,
 White N.E., Giommi P.\ 1997, IAU Circ.\ 6789
\re
 Burke B.E., Mountain R.W., Daniels P.J., Dolat V.S., 
 Cooper M.J.\ 1994, IEEE Trans.\ Nucl.\ Sci.\ 41, 375
\re
 Coe M.J., Buckley D.A.H., Charles P.A., Southwell K.A., 
 Stevens J.B.\ 1998, MNRAS 293, 43
\re 
 Coe M.J., Haigh N.J., Reig P.\ 2000, MNRAS 314, 290
\re
 Coe M.J., Orosz J.A.\ 2000, MNRAS 311, 169
\re
 Cowley A.P., Schmidtke P.C., McGrath T.K., Ponder A.L.,
 Fertig M.R., Hutchings J.B., Crampton D.
 1997, PASP 109, 21
\re
 Haberl F., Filipovi\'{c} M.D., Pietsch W., Kahabka P.\ 
 2000, A\&AS 142, 41
\re
 Haberl F., Sasaki M.\ 2000, A\&A 359, 573
\re
 Hughes J.P., Smith R.C.\ 1994, AJ 107, 1363
\re
 Israel G.L., Stella L., Angelini L., White N.E., Giommi P., Covino S.\ 
 1997, ApJ 484, L141
\re
 Lamb R.C., Prince T.A., Macomb D.J., Finger M.H.\ 
 1999, IAU Circ.\ 7081
\re
 Meyssonnier N., Azzopardi M.\ 1993, A\&AS 102, 451 (MA93)
\re
 Murphy M.T., Bessell M.S.\ 2000, MNRAS 311, 741
\re
 Ohashi T., Ebisawa K., Fukazawa Y., Hiyoshi K., 
 Horii M., Ikebe Y., Ikeda H., Inoue H.\ 
 et al.\ 1996, PASJ 48, 157
\re
 Serlemitsos P.J., Jalota L., Soong Y., Kunieda H., Tawara Y., 
 Tsusaka Y., Suzuki H., Sakima Y.\ 
 et al.\ 1995, PASJ  47, 105
\re
 Southwell K.A., Charles P.A.\ 1996, MNRAS 281, L63
\re
 Stevens J.B., Coe M.J., Buckley D.A.H.\ 1999, MNRAS 309, 421
\re
 Torii K., Kohmura T., Yokogawa J., Koyama K.\ 2000, IAU Circ.\ 7441
\re
 Tucholke H.-J., de Boer K.S., Seitter W.C.\ 1996, A\&AS 119, 91
\re
 Ueda Y., Inoue H., Ogawara Y., Fujimoto R., Yamaoka K., Kii T., 
 Gotthelf E.V.\ 1999, ISAS research Note 688
\re
 van den Bergh, S.\ 2000, PASP 112, 529
\re
 Yokogawa J., Imanishi K., Tsujimoto M., Nishiuchi M., Koyama K., 
 Nagase F., Corbet R.H.D.\ 2000, ApJS 128, 491

\end{document}